\documentstyle[12pt]{article}
\begin{document}
\normalsize
\begin{center}
{\Large A Nonlinear $sl(2)$ Dynamics and New Quasiclassical Solutions 
for a Class of Quantum Coupled Systems}\\
\vspace{7mm}
               V.P. KARASSIOV\\
{\it Lebedev Physical Institute, Leninsky prospect 53, 117924 Moscow,
Russia\\
E-mail: vkaras@sci.lpi.msk.su}
\end{center}
\begin{abstract}
Hamiltonians of a wide-spread class of strongly coupled quantum system
models are expressed as nonlinear functions of $sl(2)$ generators.
It enables us to use the $sl(2)$ formalism, in particular,
$sl(2)$ generalized coherent states (GCS)
for solving both spectral and evolution tasks. In such a manner, using 
standard variational schemes with $sl(2)$ GCS as trial functions we find
new analytical expressions for energy spectra and non-linear evolution 
equations for cluster dynamics variables in  mean-field
approximations which are beyond quasi-harmonic ones obtained earlier.
General results are illustrated on certain concrete models of quantum
optics and laser physics.
\end{abstract}

\section{Introduction}

For last decades a great attention has been paid to solve and to examine
different dynamical problems for quantum strongly coupled systems whose
interaction Hamiltonians are expressed by nonlinear functions of operators
describing subsystems (see, e.g., [1-9] and references therein). However, 
as a rule, for these purposes numerical calculations are mainly used while
analytical techniques available either deal with special forms of model
Hamiltonians (including their different semiclassical versions) and initial
quantum states [1-5,7-9] or require lengthy and tedious calculations (as it
is the case, e.g., for the algebraic Bethe ansatz [6]).

Recently, a new universal Lie-algebraic approach has been developed [10-12]
to get solutions of both spectral and evolution problems for some nonlinear 
quantum models of strongly coupled subsystems having symmetry groups 
$G_{inv}$. It was based on exploiting a formalism of polynomial Lie algebras
$g_{pd}$ as dynamic symmetry algebras $g^{DS}$ of models under study
with generators of these algebras $g_{pd}$  being $G_{inv}$-invariant 
collective (cluster) dynamic variables in whose terms model dynamics are 
described completely. (In fact, such a reformulation of original problems in 
terms of $g_{pd}$-variables is similiar to the well-known procedure of 
exclusion of cyclic variables in classical mechanics [13].) Specifically, 
this approach enabled us to develop some efficient techniques for solving 
physical tasks in the case of $g^{DS}=sl_{pd}(2)$, when model Hamiltonians 
$H$ are expressed as follows
$$ H = aV_0 +g V_+ + g^* V_- +C,\quad [V_{\alpha}, C]=0,
\quad V_- =(V_+)^+,      \eqno (1.1) $$
where $C=C([R_i])$ is a function of a set of commuting operators (model
integrals of motion) $R_i,i=1,2,...$ and $V_0, V_{\pm}$
are the $sl_{pd}(2)$ generators satisfying the commutation relations
$$ [V_0, V_{\pm}]= \pm V_{\pm}, \quad [V_-, V_+] =  \Psi (V_0+1) -
\Psi (V_0), \; \Psi (V_0)=A\prod_{i=1}^{n(\Psi)} (V_0+\lambda_i(\{R_j\})),
\eqno (1.2a)$$
$$[\Psi (R_0), V_{\alpha}]=0 \quad\forall \alpha =0,\pm, \qquad \Psi(R_0)=
\Psi(V_0)-V_+V_-, \eqno (1.2b)$$
where $n(\Psi)$ is the polynomial $\Psi$ degree in the variable
$V_0$, $\Psi(R_0)$ is the $sl_{pd}(2)$ Casimir operator (with $R_0$ being
the "lowest weight operator") and hereafter the identity operator
symbol $I$ is omitted in expressions like $\Psi (V_0 + \alpha I)$.
The structure polynomials $\Psi (V_0)$ depend additionally
on $\{R_i, i=1,\dots\}$, and their exact expressions for some wide-spread
classes of concrete models were given in [10-12] (see also Section 5).

All techniques [10-12] essentially use expansions of evolution operators
$U_{H}(t)$, generalized coherent states (GCS), energy eigenfunctions $|E_f>$
and other important physical quantities  by power series in the
$sl_{pd}(2)$ shift generators $V_{\pm}$ as well as commutation relations
(1.2)  and the characteristic equation
$$ (V_+V_- -\Psi (V_0)\equiv -\Psi (R_0))|_{L(H)}=0        \eqno (1.3)$$
fulfilled on Hilbert spaces $L(H)$  of quantum model states due to the
complementarity of groups $G_{inv}$ and algebras $sl_{pd}(2)$ [10].
Specifically, Eq. (1.3) implies a spectral decomposition
$$ L(H) =\sum_{[l_i]}\; L([l_i]), \qquad  L([l_i])\equiv
Span \{|[l_i]; v\rangle={\cal N}([l_i];v)V_+^v|[l_i]\rangle\},$$
$$ V_0|[l_i];v\rangle=(l_0 + v) |[l_i];v\rangle,
\; R_i|[l_i]; v\rangle = l_i |[l_i];v\rangle, i=0,1,...,
\; V_-|[l_i]\rangle=0          \eqno (1.4) $$
of spaces $L(H)$ in direct sums of the subspaces $L([l_i])\equiv L(l_0,l_1,
\dots)$ which are irreducible with respect to joint
actions of algebras $sl_{pd}(2)$ and symmetry groups $G_{inv}$;
lowest weights $l_0$ depend on other quantum numbers $l_i,i=1, \dots$
due to  the relation $\Psi (l_0)=0$ implied by Eq. (1.3). From the
physical point of view, the decomposition (1.4) means that the model
Hamiltonian matrices in the symmetry adapted orthonormalized bases
$\{|[l_i];v\rangle\}$ have block-diagonal forms and subspaces $L([l_i])$
describe specific "$sl_{pd}(2)$-domains" evolving independently in time
under action  of Hamiltonians (1.1). We also distinguish compact
($su_{pd}(2)$) and non-compact ($su_{pd}(1,1)$) versions of $sl_{pd}(2)$
algebras depending on whether dimensions $d([l_i])$ of the spaces $L([l_i])$
are finite or infinite.

Then, using restrictions $H_{[l_i]}$ of Eq. (1.1) on $L([l_i])$, simple
algebraic calculation schemes were developed for finding evolution operators
$U_{H}(t)=\sum_{f=-\infty}^{\infty} V_+^f \;u_f (V_0;t)$,
amplitudes $Q_v(E_f)= \langle [l_i]; v|E_f\rangle $ of energy
eigenstates $|E_f\rangle$ expansions in orthonormalized
bases $\{|[l_i]; v\rangle \}$) and appropriate energy
spectra $\{E_f\}$ of bound states [10]. In the paper [11] some explicit 
integral representations were found for amplitudes $Q_v(E)$, eigenenergies 
$\{E_a\}$ and "evolution coefficients" $u_f(V_0;t)$  with the
help of a specific "dressing" (mapping) of solutions of some auxiliary
exactly solvable tasks with the dynamic algebra $sl(2)$.

However, all exact results obtained do not yield simple working formulas
for analysing models (1.1) and revealing different physical effects (e.g.,
a structure of collapses and revivals of the Rabi oscillations [2,8],
bifurcations and singularities of quasiclassical solutions [5] etc.) at
arbitrary initial quantum states of models. Therefore, it is necessary to
develop some simple techniques, in particular, to get some closed, perhaps,
approximate expressions for evolution operators, energy eigenvalues and wave
eigenfunctions, which would describe main physical peculiarities of
model dynamics with a good accuracy (cf. [5,8,9]). Below we examine some
possibilities along these lines for models (1.1)-(1.4) by means
of reformulating them in terms of the usual $sl(2)$ algebra formalism
and developing variational schemes corresponding to quasiclassical
approximations (QAs) for these models by analogy
with developments [5,14-16].

The work is organized as follows. In Section 2 we first reformulate models
(1.1)-(1.4) in terms of the usual Lie algebra $sl(2)$ formalism, and then
discuss possibilities of extending the standard $sl(2)$-techniques
to analyse such reformulated models. In Section 3 a scheme
is given for obtaining QAs of these models by using
variational principles [5,17] and energy functionals
constructed with the help of the $SL(2)$ group GCS [16]; these QAs are new 
for original models because they take into account a strong coupling of 
interacting subsystems in contrast with standard QAs. Specifically, in 
such a manner new analytical expressions are obtained for energy spectra 
which are essentially non-equidistant on each subspace $L([l_i])$ with its 
dimension $d([l_i])\geq 4$. In Section 4 we discuss such approximations for
a quasiclassical description of dynamics of $sl(2)$-clusters (characteristic
model exicitations) and time evolution of uncoupled dynamical variables;
specifically, nonlinear evolution equations of the Bloch type
are obtained for
$sl(2)$-cluster variables. In Section 5 a specification of general results
is given for a class of models widely used in quantum optics and laser
physics. In conclusion some of prospects of developing this approach are
discussed.
 
\section {A nonlinear $sl(2)$ formulation and a general operator analysis
of quantum models with linear $sl_{pd}(2)$ dynamic algebras }

We can reformulate  models (1.1), (1.4) in terms of $sl(2)$
generators using a realization of the $sl_{pd}(2)$ algebras in terms of
special elements of extended enveloping algebras ${\cal U}_{\Psi}(sl(2))$ of
the familiar algebra $sl(2)$ [12]. This realization is established via the
generalized Holstein-Primakoff mapping [10]
$$Y_0 = V_0-R_0\mp \hat J,\; Y_+= V_+ [\Phi (Y_0)]^{-1/2},\;
\Phi (Y_0)=\frac{\Psi (Y_0+R_0\pm\hat J+1)}{(\hat J
\mp Y_0)(\pm\hat J+1+Y_0)}, \;Y_-=(Y_+)^+, \eqno (2.1a)$$
$$ [Y_0, Y_{\pm}]= \pm Y_{\pm}, \quad [Y_-, Y_+] = \mp 2 Y_0 \eqno (2.1b)$$
where $Y_{\alpha}$ are the $sl(2)$ generators, $R_0, \mp\hat J$ are lowest
weight operators of the $sl_{pd}(2)$ and $sl(2)$ algebras respectively:$\,
\Psi (R_0))|_{L(H)}=0, [\hat J(\hat J\pm1)\mp Y_+Y_- -Y_0^{(2)}]|_{L(H)}=0$.
Appropriate specifications of Eqs. (2.1a) on
subspaces $L([l_i])$ are obtained
by the substitution $R_0\rightarrow l_0,\hat J\rightarrow J$ and hereafter
upper/lower signs in (2.1) corresponding to the $su(2)$/$su(1,1)$ algebras
are chosen for finite/infinite dimensions $d([l_i])$ of
the spaces $L([l_i])$. Note that, by definition (2.1),
functions $\Phi (Y_0)$ on $L([l_i])$ are
polynomials of the $(n(\Psi)-2)$-th degree in the variable $Y_0$ at relevant
values of $J$ [10,12].

Then, using Eqs. (2.1) one may  re-write Hamiltonians (1.1) in terms of
$Y_{\alpha}$ as follows,
$$ H = aY_0 + Y_+ g(Y_0) + g^+(Y_0) Y_- + C', $$
$$g(Y_0)= g\sqrt{\Phi(Y_0)},\qquad C'\equiv C'([R_i],\hat J)=
C([R_i])+a(R_0\pm\hat J)   \eqno (2.2) $$
Restrictions $H_{[l_i]}\equiv P_{[l_i]} H$ of Hamiltonians (1.1) on spaces
$L([l_i])$ (with $P_{[l_i]}=\sum_{v}\,|[l_i];v\rangle\langle v;[l_i]|$ being
appropriate central projectors) are obtained by the substitution $R_0
\rightarrow l_0,\hat J\rightarrow J$ in Eq.(2.2). Respectively, basis
vectors $|[l_i];v\rangle$ of spaces $L([l_i])$ are given in terms of
$Y_{\alpha}$ as follows,
$$|[l_i];v\rangle= {\cal N}(J,v)(Y_+)^v|[l_i]\rangle \eqno (2.3)$$
where ${\cal N}^{-2}(J,v)=v!(2J)!/(2J-v)!\; \mbox{for}\; su(2)\;\mbox{and}\;
{\cal N}^{-2}(J,v)=v!\Gamma (2J+v)/\Gamma (2J)\;\mbox{for}\; su(1,1)$.
Evidently, Eq. (2.2) resembles Hamiltonians of semi-classical $sl(2)$
"linearized" versions of matter-radiation interaction models [4,8,9,12] but
with operator (intensity-dependent) coupling coefficients $g(Y_0)$
(cf. [3,4,7]). Emphasize, however, a collective (not associated
with a single subsystem) nature of operators $Y_{\alpha}$ that
leads, when substituting $g(Y_0)$
in Eq. (2.2) by an "effective coupling constant", to a non-standard
("cluster") QA of original models [10] distinguished from standard
semi-classical limits [8,18,19] where a part of interacting subsystems is
described classically.

If $n(\Psi)=2$, then $\Phi(Y_0)=1, sl_{pd}(2)=sl(2), R_0=\mp\hat J $, and we
have a powerful tool for solving both spectral and evolution tasks yielded
by the GCS formalism [16] related to the $SL(2)$ group
displacement operators
$$S_Y(\xi=re^{i \theta})=\exp(\xi Y_+-\xi^* Y_-)=\exp[t(r)e^{i \theta} Y_+]
\exp[-2\ln c(r) Y_0]\exp[-t(r)e^{-i\theta} Y_-]=$$
$$\sum_{f=-\infty}^{\infty} Y_+^f S^Y_f(Y_0;\xi), \quad Y_+^{-k}
\equiv Y_-^k \left ([\Psi_2 (Y_0)]^{(k)}\right )^{-1}
\;\mbox{for}\; k>0 \eqno (2.4a)$$
where $t(r)= \tan r/\tanh r, c(r)=\cos r/\cosh r, s(r) =
\sin r /\sinh r$ for $su(2)/su(1,1)$ and
$$[\Psi_2 (Y_0)]^{(k)}\equiv  (\pm 1)^k(\pm\hat J+Y_0)^{(k)}
(\pm\hat J-Y_0+k)^{(k)}, \quad  A^{(x)}\equiv A(A-1)...(A-x+1)
\eqno (2.4b)$$
$$S^Y_f(Y_0;\xi)= \frac{(e^{i\theta}t(r))^f}{f!}{_2F_1 (\mp J-Y_0,
-Y_0\pm\hat J +1;f+1; \pm [s(r)]^2 )}\exp[-2\ln c(r) Y_0]
\eqno (2.4c)$$
with ${_2F_1(...)}$ being the Gauss hypergeometric function [20].

Specifically, in this case, using the well-known $sl(2)$ transformation
properties of operators $Y_{\alpha}$ under the action of $S_Y(\xi)$ [16,12],
$$S_Y(\xi)Y_{+}S_Y(\xi)^{\dagger}\equiv Y_{+}(\xi)=[c(r)]^2 Y_{+}
\pm e^{-i\theta} [ s(2r) Y_0 - e^{-i\theta} [s(r)]^2 Y_-],$$
$$ S_Y(\xi)Y_{0}S_Y(\xi)^{\dagger}\equiv Y_{0}(\xi)=c(2r) Y_{0}
- \frac{s(2r)}{2} [ e^{i\theta} Y_{+} +  e^{-i\theta} Y_{-},
 \; Y_{-}(\xi)=(Y_{+}(\xi))^{\dagger},  \eqno (2.5)$$
Hamiltonians $H$ can be transformed into the form
$$ \tilde{H}(\xi)=S_Y(\xi)HS_Y(\xi)^{\dagger}=C' +Y_0 A_0(a, g; \xi)+
Y_+A_+(a, g; \xi) +Y_- A^*_+(a, g; \xi)  \eqno (2.6a)$$
At the values $\xi_0=\frac{g}{|g|} r$ of the parameter $\xi$ with $\tan 2r/
\tanh 2r= \frac{2|g|}{a}$ for $su(2)/su(1,1)$ one gets $A_+(a, g; \xi)=0$,
and the Hamiltonian $\tilde{H}_{[l_i]}(\xi)$ takes the form
$$\tilde{H}(\xi_0)= C' +Y_0\sqrt{a^2\pm 4 |g|^2}
\eqno (2.6b)$$
which is diagonal on eigenfunctions $|[l_i];v\rangle$. Therefore, original
Hamiltonians $H$ have within each $L([l_i])$ equidistant spectra with
eigenenergies
$$E([l_i];v)= \tilde C +(\mp J+v)\sqrt{a^2\pm 4 |g|^2}, \quad \tilde C=
C'([l_i];J)      \eqno (2.7a)$$
and eigenfunctions
$$|[l_i];v;\xi_0\rangle= S_Y(\xi_0)^{\dagger}|[l_i];v \rangle=
\exp(-\xi_0 Y_++ \xi_0^*Y_-)|[l_i];v; \rangle=
 \sum_{f\geq 0} S_{f v}(J;g,r) |[l_i];f\rangle,$$
$$S_{f v}(J;g,r)=
\frac{[c(r)]^{2(\pm J-v)}(-\frac{g}{|g|}t (r))^{f-v}{\cal N}(J,v)}{(f-v)!
{\cal N}(J,f)}{_2F_1(-v,-v\pm 2J+1;f-v+1; \pm [s(r)]^2)},   \eqno(2.7b)$$
where  $t(r)=\pm\left (-a +\sqrt{a^2\pm 4|g|^2}~ \right )/2|g|$ and
${\cal N}(J,...)$ are normalization constants from Eq. (2.3).

Similarly, if $sl_{pd}(2)=sl(2)$, operators $S_Y(\xi (t))$ are "principal"
parts in the evolution operators $U_{H}(t)= \exp(i\alpha (t) Y_0)
S_Y(\xi (t))$ with $\alpha (t), \xi (t)$ being $c$-number functions in
$t$ which are determined from disentangling the exponent $\exp(\frac{i t}
{\hbar}H)$ (or, when $g$ are time-dependent functions, from a set of
non-linear differential equations corresponding to
classical motions) [16,19].

However, for arbitrary degrees $n$ of polynomials $\Psi (V_0)$ Hamiltonians
(2.2) are essentially nonlinear in $sl(2)$ generators $Y_{\alpha}$, and,
therefore, the situation is very changed. Specifically, in general cases it
is unlikely to diagonalize $H$ with the help of operators $S_Y(\xi)$ since
analogs of Eq. (2.6a),
$$\tilde{H}(\xi)=S_Y(\xi)HS_Y(\xi)^{\dagger}= aY_0 (\xi) +
Y_+(\xi) g(Y_0(\xi)) +  g^+(Y_0(\xi)) Y_-(\xi) +C,'
    \eqno (2.8) $$
and even their restrictions $\tilde{H}_{[l_i]}(\xi)=S_Y(\xi)H_{[l_i]}
S_Y(\xi)^{\dagger}$ on multi-dimensional spaces $L([l_i])$ contain (after
expanding $g(Y_0(\xi))$ in power series) many terms with higher powers of
$Y_{\pm}$. The task is also not simplified when using in Eq. (2.8) 
operators $S_V(\xi)=\exp(\xi V_+-\xi^* V_-)$ instead of $S_Y(\xi)$ because
we have not suitable analogs of the "disentangling theorem" (2.4) and
finite-dimensional transformations (2.5) for operators $S_V(\xi)$ [12].
Therefore it is necessary to use in Eq. (2.8) more general (perhaps,
non- or multi-parametric) forms of diagonalizing operators $S$ given, e.g.,
by power series
$$S=\sum_{f=-\infty}^{\infty} Y_+^f S_f(Y_0),  \quad Y_+^{-k}\equiv
 Y_-^k \left ([\Psi_2(Y_0)]^{(k)}\right )^{-1} \; \forall\; k>0
    \eqno (2.9)$$
with undetermined (unlike Eq. (2.4)) coefficients $S_f(Y_0)$ and satisfying
the unitarity conditions $S S^\dagger =S^\dagger S=I$.

Substituting Eq. (2.9) in the scheme (2.8) one gets after some algebra
nonlinear analogs of Eqs. (2.6a)
$$\tilde{H}=S H S^{\dagger}= C'+\sum_{f=-\infty}^{\infty}Y_+^f
\tilde{h}_f(Y_0)  \eqno (2.10a)$$
where
$$ \tilde{h}_f(Y_0)= \sum_{k=-\infty}^{\infty} [\Psi_2 (Y_0)]^{(k)}
S^*_k(Y_0-k) [a (Y_0-k) S_{k+f}(Y_0-k) +$$
$$g\sqrt{\Phi(Y_0-k)}S_{k-1+f}(Y_0-k+1)+ g^*\sqrt{\Phi(Y_0-k-1)}
S_{k+1+f}(Y_0-k-1) \Psi_2(Y_0-k)]   \eqno (2.10b)$$
and $\tilde{h}_{-f}(Y_0)=\tilde{h}^*_f(Y_0-f)[\Psi_2(Y_0)]^{(f)},
\;[\Psi_2(Y_0)]^{(-f)}\equiv([\Psi_2(Y_0+f)]^{(f)})^{-1}$ for all\, $f>0$.

The conditions $\tilde{h}_f(Y_0)=0$ for all $f\neq 0$ yield nonlinear 
analogs of Eqs. (2.6b), (2.7b),
$$\tilde{H}^0= S^0 H S^{0 \dagger}=C' +\tilde{h}^0_0(Y_0)=
\sum_{k=-\infty}^{\infty} [\Psi_2 (Y_0)]^{(k)}
S^{0*}_k(Y_0-k) [a (Y_0-k) S^0_{k}(Y_0-k) +$$
$$g\sqrt{\Phi(Y_0-k)}S^0_{k-1}(Y_0-k+1)+ g^*\sqrt{\Phi(Y_0-k-1)}
S^0_{k+1}(Y_0-k-1) \Psi_2(Y_0-k)],     \eqno (2.11a)$$
$$ E([l_i];v)= \tilde C+\langle[l_i];v|\tilde{h}^0_0(Y_0)|[l_i];v\rangle =
\tilde C + \tilde{h}^0_0(\mp J+v)      \eqno (2.11b)$$
expressed in terms of the coefficients $S^0_f(Y_0)$ which satisfy Eqs.
(2.10b) with $\tilde{h}_{f}(Y_0)=0$ for $f\neq 0$ and simultaneuosly
are solutions of the set of algebraic operator equations,
$$[-a Y_0 +\tilde{h}^0_0(Y_0+f)]S^0_{f}(Y_0)=$$
$$g\sqrt{\Phi(Y_0)}S^0_{f-1}(Y_0+1)+ g^*\sqrt{\Phi(Y_0-1)}
S^0_{f+1}(Y_0-1) \Psi_2(Y_0)\quad\forall\, |f|\,>1
    \eqno (2.12)$$
resulting from the condition $S^0 H=[C'+\tilde{h}^0_0(Y_0)]S^0$ [12] where
$S^0$ is given by Eq. (2.9) with "coefficients" $S^0_{f}(Y_0)$.

In the case of $\Phi(Y_0)=1$ Eqs. (2.12) are solved in terms of
hypergeometric functions ${_2F_1(...)}$ as it follows from
Eq. (2.4c), but in general they,
probably, determine certain $q$-special functions due to relations of
$sl_{pd}(2)$ algebras with certain $q$-deformed algebras [21]. Without using
any specifications of operators $S$, due to the relation $Q_v(E_f)=
S_{f-v}^*(\pm J+v)\frac{{\cal N}(J,v)}{{\cal N}(J,f)}$, the task of solving
these equations is equivalent to that for finding amplitudes $Q_v(E_f)$
related to new classes of orthogonal functioms [10]. Note that this task
is simplified in the compact ($su(2)$) case, when all subspaces $L([l_i])$
have finite dimensions $d([l_i])=2J+1$, and all series in Eqs. (2.9)-(2.11)
are terminating due to Eq. (1.3) and the relation
$(Y_{\pm})^{2J+1}|_{L([l_i])} =0$. Therefore, eigenfunctions
$|E_v \rangle=S^{\dagger}|[l_i];v\rangle$ may be represented by polynomials
$$|E_v \rangle=\sum_{f=0}^{2J} Q_f^v Y_+^f |[l_i] \rangle=
A_v\prod_r (Y_+ - \kappa _r^v) |[l_i] \rangle,
\eqno (2.13)$$
where amplitudes $Q_f^v$ are expressed as symmetric functions in variables
$\kappa_r^v$:
$$Q_{2J}^v=A_v,\qquad Q_{2J-1}^v=-A_v\sum_{r=0}^{2J}\kappa _r^v, ...,$$
$$Q_{2J-f}^v=(-1)^fA_v\sum_{1\leq r_1<r_2<...<r_f\leq 2J}\kappa _{r_1}^v
\kappa _{r_2}^v ...\kappa _{r_f}^v , ...,\quad Q_0^v=
(-1)^{2J} A_v\kappa _{r_1}^v\kappa _{r_2}^v ...\kappa _{r_{2J}}^v,
\eqno (2.14)$$
At the same time eigenenergies $E([l_i];v)$ determined by the boundary
condition [10]
$$[(l_0 +2J)a- E([l_i];v)+C([l_i])]Q_{2J}^v + g Q_{2J-1}^v=0, \eqno (2.15)$$
can be written down in the form
$$E([l_i];v)=\tilde C+Ja -g\sum_{r=0}^{2J}\kappa _r^v, \quad \tilde C=
C([l_i])+(l_0 +J)a                                       \eqno (2.16)$$
of a sum of $2J+1$ spectral functions as it is prescribed by the
algebraic Bethe ansatz [6]. In fact, Eqs. (2.13)-(2.16) give for models given
by Eqs. (1.1)-(1.4) a new, $G_{inv}$-invariant formulation of this ansatz
in terms of the $su(2)$ algebra which is simpler and more efficient
in comparison with its initial non-invariant version [6] because
the algorithm [10] for finding amplitudes $Q_f^v$ and eigenenergies
$E([l_i];v)$  does not require a preliminary determination of parameters
$\kappa _r^v$. We also note that in the resonance case (when $a=0$
in (2.2)), using Eqs. (2.12), one can get analytical solutions for
amplitudes $Q_f^v$ in the form of multiple sums which, however, are
not suitable for practical purposes.

So,  direct generalizations of "linear" schemes (2.6) to the case of
non-linear Hamiltonians (2.2) do not yield simple analytical formulas for
exact solutions of spectral tasks; a similar situation is also with
respect to evolution problems. Nevertheless, the formalism of the $SL(2)$
GCS $|[l_i];v;\xi\rangle= S_Y(\xi)^{\dagger}|[l_i];v \rangle$ can be
an efficient tool for analysing such models [5,10,14-16] and for getting
approximate analytical solutions of both spectral and evolution problems.
Specifically, a simplest example of such approximations was given in [10]
by mapping (with the help of the change $V_{\alpha}\rightarrow Y_{\alpha}$)
Hamiltonians (1.1) into Hamiltonians $H_{sl(2)}$ which are linear in $sl(2)$
generators $Y_{\alpha}$ (but with modified constants $\tilde a, \tilde g$)
and have on each fixed subspace $L([l_i])$ equidistant energy spectra
given by Eq. (2.7a). However, this (quasi)equidistant approximation,
in fact, corresponding to a substitution of certain effective coupling
constants $\tilde g$ instead of true operator entities $g(Y_0)$ in
Eq. (2.2), does not enable to display many peculiarities of models (1.1)
related to essentially non-equidistant parts of their spectra. Therefore,
it is needed in corrections, e.g., with the help of iterative schemes
[8,14,15]; specifically, one may develop perturbative schemes by using
expansions of operator entities $g(Y_0)$ in Taylor series in
$Y_0$ as it was made implicitly for the Dicke model in [8,9]. But there
exists a more effective, incorporating many peculiarities of models (1.1),
way to amend the quasi-equidistant approximation.

\section{$SL(2)$ quasiclassical approximations: energy functionals and
variational energy spectra}

This way is in applying $SL(2)$ GCS $|[l_i];v;\xi\rangle$ from Eq. (2.7b) as
trial functions in the variational schemes [17] of determining energy
spectra and quasiclassical dynamics [5,15]. Indeed, because of
the isomorhism of quantum and quasiclassical dynamics for $sl(2)$
linear Hamiltonians [15,16],
the results (2.7) can be obtained with the help of the variational scheme
determined by the stationarity conditions
$$a)\;\frac{\partial {\cal H}([l_i];v;\xi)}{\partial \theta}=0,\qquad
b)\;\frac{\partial {\cal H}([l_i];v;\xi)}{\partial r}=0
\eqno (3.1)$$
for the energy functional ${\cal H}([l_i];v;\xi)=\langle[l_i];v;\xi|H
|[l_i];v;\xi\rangle= \langle[l_i];v|C' +Y_0 A_0(a, g; \xi)+
Y_+ A_+(a, g; \xi) +Y_- A^*_+(a, g; \xi) |[l_i];v\rangle $ (cf. (2.6a)).
Similarly, following the standard variational approach [17,5], the
calculation schemes (3.1) may be extended to the case of nonlinear
Hamiltonians (2.2) by using the energy functional
$${\cal H}^{cq}([l_i];v;\xi)=\langle[l_i];v;\xi|H|[l_i];v;\xi\rangle=
\langle v;[l_i]|\tilde{H}(\xi)|[l_i];v\rangle      \eqno (3.2)$$
where superscript $cq$ denotes "cluster" (strongly correlated) QAs
(as contrasted with standard QAs dealing with weakly or non-correlated
subsystems) and $\tilde{H}(\xi)$ are given by Eq. (2.8) or Eqs.
(2.10) with $S_{f}(Y_0)=S^Y_f(Y_0;\xi)$ from Eqs. (2.4c).

Note that, due to the Hermitian conjugacy relations $Y_+ g(Y_0)=
(g^+(Y_0) Y_-)^{\dagger}$, the condition (3.1a) gives $e^{i\theta}=g/|g|$
(as in the linear case); in fact, this condition (3.1a) can be eliminated at
once by the simple gauge transformation $Y_{\alpha}\rightarrow
\exp (-i\alpha\gamma) Y_{\alpha}, g=|g|\exp (i\gamma),$ in Eqs.(2.2) which
preserves commutation relations (2.1b). Furthermore, due to the
form of trial functions and the unitarity of operators $S_Y(\xi)$, it is
sufficiently to solve Eq. (3.1b) only for finding states $|[l_i]; v=0;
\xi\rangle$ and to use the only real root of one of Eqs. (3.1b)  for all
$v$ that ensures automatically the orthogonality of eigenfunctions.
Naturally, results thus
obtained are not expected to coincide with exact solutions on all subspaces
$L([l_i])$ due to an essential non-linearity of Hamiltonians (2.2b) and
their non-equivalence (unlike Eq. (2.6b)) to diagonal parts of Eq. (2.10a);
however, they yield "smooth" (analytical) solutions which are
in a sense most close to exact ones (cf. [5,14]).
Without discussing all aspects of such extensions we give below two
approximations for energy spectra obtained by inserting in Eq. (3.2)
$\tilde{H}(\xi)$ given by Eqs. (2.10) and Eq. (2.8) respectively.

In the first case, using Eq. (2.5) for $Y_{0}(\xi_0)$, Eq. (2.7b) for
$|[l_i];v;\xi\rangle$ and defining relations for the $sl(2)$ algebra, one
gets the following "cluster" QAs
$$E^{cq}([l_i];v)=\tilde C+a(v\mp J) c(2r)+
\Re\{g\langle[l_i];v;\xi_0|Y_+\sqrt{\Phi (Y_0)}|[l_i];v;\xi_0\rangle\}=$$
$$\tilde C+a(v\mp J) c(2r)-
2|g|\sum_{f\geq 0} |S_{f v}(J;g,r) S_{f+1 v}(J;g,r)| \sqrt{\Psi(l_0+1+f)},
   \eqno (3.3a)$$
$$E^{cq}([l_i];0)=\tilde C \mp a J c(2r)-2|g|[c(r)]^{\pm 4J}
\sum_{f\geq 0}\frac{\sqrt{\Phi(\mp J+f)} (t(r))^{2f+1}}{f!(f+1)!
{\cal N}^2(J,f+1)}   \eqno (3.3b)$$
for energy eigenvalues $E^{cq}([l_i];v) ={\cal H}^{cq}([l_i];v;\xi_0)$
where $\tilde C=C([l_i])+a(l_0\pm J), \,\xi_0=rg/|g|,\,
2\Re\{A\}=A^*+A, c(r)=\cos r/\cosh r, \Psi (l_0+1+f)=\Phi (\mp J+f)
(2J\mp f)(f+1)$, ${\cal N}(J,...)$ are normalization constants from Eq.
(2.3) and functions $S_{fv}(J;g,r)$ are given by Eq. (2.7b) but with values
of the parameter $r$ determined by real solutions of the algebraic equation
$$\frac{2aJ}{|g|}\alpha(1\pm\alpha^2)^{\pm 2J-1} = \sum_{f\geq 0}
\frac{\sqrt{\Phi(\mp J+f)} (t(r))^{2f}}{f!(f+1)!N^2(J,f+1)}
[\mp 4\alpha^2 j +(1\pm\alpha^2)(2f+1)],$$
$$ \alpha =t(r)=\tan r/\tanh r \eqno (3.4)$$
which follows from Eqs. (3.1b) and (3.3b). Obviously, unlike the linear case,
diagonalizing values of $r$ depend on both constants $g, a$ and quantum
numbers $l_i$ labeling $G$-invariant subspaces $L([l_i])$.

In the second case it is difficult to obtain exact analytical formulas
like Eq. (3.3) due to presence of square roots in Eq. (2.8). However we can
get another approximation for energy spectra if replacing the energy
functionals (3.2) by their (corresponding to the Ehrenfest theorem with
respect to cluster variables $Y_i$) mean-field approximations
$${\cal H}^{cmf}([l_i];v;\xi) =
a<Y_0(\xi)> + 2\Re\{<Y_+(\xi)>\tilde g(<Y_0(\xi)>)\}+\tilde C,$$
$$ <Y_{\alpha}(\xi)> =\langle v;[l_i]|Y_{\alpha}(\xi)|[l_i];v\rangle,
\quad 2\Re\{A\}= A+A^*
   \eqno (3.5) $$
Then, inserting Eqs. (2.5) in Eq.(3.5), one finds the "cluster" mean-field
approximations $E^{cmf}([l_i];v)$ for eigenenergies,
$$E^{cmf} ([l_i];v)= \tilde C+a(v\mp J) c(2r)-
2|g|(J\mp v)s(2r) \sqrt{\Phi((\mp J+v)c(2r))}       \eqno (3.6a)$$
where $r$ is determined from the equation
$$\frac{a}{|g|}s(2r)= \pm 2c(2r)\sqrt{\Phi(\mp J c(2r))}+
\frac{J[s(2r)]^2 \Phi '(\mp J c(2r))}{\sqrt{\Phi(\mp J c(2r))}},
\;\Phi '(\mp J c(2r))=\frac{\partial\Phi(x)}{\partial x}|_{x=\mp J c(2r)}
     \eqno (3.6b)$$
Let us make some remarks concerning results obtained.

{\it Remark 1.} As is seen from Eq. (3.3), its general structure coincides
with the energy formula given by Eq. (2.16), and spectral functions
$|S_{f v}(J;g,r) S_{f+1 v}(J;g,r)| \sqrt{\Psi(l_0+1+f)} = E^{\Phi}_f(r;J;v)$
are nonlinear in the discrete variable $v$ labeling energy levels within
$L([l_i])$ that provides a non-equdistant character of energy spectra within
fixed subspaces $L([l_i])$ at $d([l_i])>3$. Besides, due to the presence of
square roots in Eqs. (3.3), (3.6a) different eigenfrequencies
$\omega_v\equiv E([l_i];v)/\hbar$ are incommensurable: $\; m\omega_{v_1}
\neq n\omega_{v_2}$ that is an indicator of complex dynamics manifesting
in such phenomena as collapses-revivals of the Rabi oscillations [2,8] and
singular and pre-chaotic dynamic regimes in phase spaces of
models [13,22,23]. Evidently, it is hardly possible to obtain
these features of models by using GCS related to uncoupled subsystems
(cf. [4,8,18] and Section 5 of the present paper).

{\it Remark 2.} In the compact ($su(2)$) case the r.h.s. of Eq. (3.4) is a
polynomial of the degree $2J+1$, and, in general, Eq. (3.4) may have
$2J+1$ different roots $r_i$ corresponding to $2J+1$ different stationary
values of the energy functional ${\cal H}([l_i];v;\xi)$. Therefore, one may
assume that it is possible to get more simple expressions for certain
$E([l_i];v)$  using $E([l_i];0)$ from Eq. (3.3b) with different real roots
$r_i$ of Eq. (3.4); specifically, it is the case for dimensions $d([l_i])=2$
when Eqs. (3.3)-(3.4) give exact results. However, using the well-known
expressions for overlap integrals of $SU(2)$ GCS [16], one can show that
in general only two $SU(2)$ GCS with different real roots $r_i$ may be
mutually orthogonal.

{\it Remark 3.} Obviously, Eq. (3.3) generalizes Eq. (2.7a) for the
(quasi)equidistant approximation abovementioned. Indeed, when replacing the
functions $\Phi (\mp J+f)$ by their certain (and the same for all labels $v$)
"average" values, series in (3.3), (3.4) are summed up, and Eq. (3.3) is
reduced to Eq. (2.7a); Taylor series expansions of functions $\sqrt{\Phi
(\mp J+f)}$ provide perturbative corrections related to higher degrees of the
an-harmonicity of Hamiltonians (2.2). At the same time Eqs. (3.5)-(3.6) yield
an intermediate (related to a more fine "averaging" procedure (3.5)) 
approximation retaining the main characteristic feature of Eq. (3.3) (a 
non-equdistant character of energy spectra within fixed subspaces $L([l_i])$) 
but being simpler in its form that is important from the practical point of
view. Besides, Eqs. (3.4) and (3.6b) are simplified in the resonance case
when $a=0$.

{\it Remark 4.} In fact, solving Eqs. (3.1b) one  can get a whole series of
competitive potential solutions (corresponding to different roots $r_i$ and
$v$ ) which may approximate
exact ones with a good accuracy in particular parts of energy spectra. (This
situation resembles that occuring in the stationary phase calculations
of the path integral approach when one needs to
take into account contributions of several classical trajectories [24,25].)
A final selection of the most adequate value $r_0$ may be made with the help
of a "quality criterion" of QCAs on subspaces $L([l_i])$. For example, one
can estimate an accuracy of QCAs obtained by means of the "energy error"
functionals [10]
$$\delta_{[l_i]}^p(H,H^{cq/cmf})= |\mathop{Tr}~_{[l_i]}(H-H^{cq/cmf})^p|/
|\mathop{Tr}~_{[l_i]}(H)^p|, \quad  p=1,2      \eqno (3.7) $$
giving "energy-trace" proximity meausures of the exact Hamiltonians (2.4)
and their QCAs
$$H^{cq}(\{Y_i\};\xi_0)=\sum_{[l_i], v}E^{cq}([l_i];v) |[l_i];v;\xi_0\rangle
\langle \xi_0;v;[l_i]|=\tilde C+S_Y(\xi_0)^+\tilde{h}_f(Y_0;\xi_0)S_Y(\xi_0),
\eqno (3.8a)$$
$$H^{cmf}(\{Y_i\};\xi_0)=\sum_{[l_i], v}E^{cmf}([l_i];v)|[l_i];v;\xi_0\rangle
\langle \xi_0;v;[l_i]|=$$
$$ \tilde C+S_Y(\xi_0)^+[a(Y_0) c(2r)\pm 2|g| Y_0 s(2r)
[\phi_{n-2}((Y_0)c(2r))]^{1/2}]S_Y(\xi_0)  \eqno (3.8b)$$
on subspaces $L([l_i])$; $\mathop{Tr}~_{[l_i]} A =\sum_{v}
\langle v;[l_i]|A|[l_i];v\rangle$. Furthermore, functionals (3.7) may be used
in alternative "minimization schemes" of determining the paprameter $r_0$.

\section{Variational quasiclassical dynamics of $SL(2)$-clusters
and time evolution of uncoupled variables}

The energy functionals (3.2) and their mean-field approximations (3.5) may be
also used for a quasiclassical analysis of time
evolution of cluster dynamical
variables related to the $sl(2)$ generators $Y_i$ (cf. [5]). As is known,
when Hamiltonians (2.2) are linear in $sl(2)$ generators,
quasiclassical dynamics is isomorphic to the exact quantum one [14-16] and is
described by the classical Hamiltonian equations [5,14,16]
$$ \dot q =\frac{\partial {\cal H}}{\partial p}, \qquad \dot p =
-\frac{\partial {\cal H}}{\partial q}, \qquad {\cal H}=
\langle z(t);[l_i]|H|[l_i];z(t)\rangle \eqno (4.1a)$$
for "motion" of the canonical parameters $p, q$ of the $SL(2)$ GCS
$|[l_i];z(t)\rangle=\exp(-z(t) Y_++ z(t)^*Y_-)|[l_i]\rangle$ as trial
functions in the time-dependent Hartree-Fock variational scheme [17] with the
Lagrangian ${\cal L}=\langle z(t);[l_i]|(i\partial/\partial t -H)|[l_i];
z(t)\rangle;\, q=\theta,\, p=\langle z(t);[l_i]|Y_0|[l_i];z(t)\rangle
=\mp J c(2r),\, z =r\exp (-i\theta)$. An equivalent formulation can be given
in ${\bf Y}=(Y_1,Y_2,Y_0)$ space using $sl(2)$ vector Euler-Lagrange
equations [5],
$$ \dot {\bf y}=\frac{1}{2} {\bf \bigtriangledown} {\cal H}\times
{\bf \bigtriangledown}{\cal C}, \quad {\bf y}= (y_1,y_2,y_0),
\; y_i =\langle z(t);[l_i]|Y_i|[l_i];z(t)\rangle,
\quad {\bf \bigtriangledown} =(\partial/\partial y_1,
\partial/\partial y_2,\partial/\partial y_0), $$
$${\cal C}=\pm y_0^2+y_1^2+y_2^2=\pm J^2, \; y_{\pm}=y_1\pm i y_2,
\;{\bf A}\times{\bf B}= (A_2B_0-A_0B_2,A_0B_1-A_1B_0,A_1B_2-A_2B_1)
            \eqno (4.1b)$$
which yield linear quasiclassical Bloch-type equations for $sl(2)$ linear
Hamiltonians [19].
 
In the general case of nonlinear Hamiltonians (2.2) Eqs.
(4.1) with ${\cal H}$ given by Eqs. (3.2) and (3.5) at $\xi=z^*, v=0$
also describe a quasiclassical $SL(2)$ "cluster" dynamics of models
under study which, however, is not isomorphic to the exact quantum one [5].
Besides, Eqs. (4.1) obtained with the
help of GCS $|[l_i];z(t)\rangle$ describe dynamics of $SL(2)$ clusters within
each subspace $L([l_i])$ separately. Specifically, Eqs. (4.1a) determine a
"$sl(2)$ linearized" QA $U^l_{H;[l_i]}(t)\propto P_{[l_i]}
\exp(-z(t) Y_++ z(t)^*Y_-)$ of $L([l_i])$-restricted evolution operators
$U_{H;[l_i]}(t)=P_{[l_i]}U_{H}(t)$ when initial wave functions
$|\psi_0\rangle$ are equal to $|[l_i]\rangle$ (cf. [19]); in a sense, this
approximation is equivalent to that obtained by substitutions in Eq. (2.2b)
time-dependent coupling functions $g(t)$ (compatible with solutions of
Eqs. (4.1a)) instead of $g(Y_0)$ (cf. [4,23]). However, for general
initial wave functions $|\psi_0\rangle\in L(H)$ it is necessary to generalize
these equations, e.g., by using GCS $\exp(-z(t)Y_++ z(t)^*Y_-)|\psi_0\rangle$.
Without dwelling on a detailed analysis of this topic we write down examples
of Eqs. (4.1) when appropriate explicit expressions for ${\cal H}$ are
obtained from Eqs. (3.3b) and (3.6a) at $v=0$ by means of the substitutions
$$2|g|\rightarrow g e^{-i q}+ g^* e^{i q}= 2[\Re\{g\}\cos q +
\Im\{g\}\sin q],\; 2i\Im\{g\}=g-g^*,\;2\Re\{g\}=g+g^*, $$
$$\mp J c(2r) =y_0=p,\quad \cos q = \frac{- y_1}{\sqrt{\pm (J^2-y_0^2)}},
\quad \sin q = \frac{ y_2}{\sqrt{\pm (J^2-y_0^2)}}      \eqno (4.2)$$
where the first line is taken from the substitution $\xi_0=rg/|g|\rightarrow
z^*= r\exp(i q)$ in Eqs.(3.3) and the second one is a direct consequence of
Eqs. (2.5).

Then, from Eqs. (3.3b), (4.1b) and (4.2) one gets essentially nonlinear
quasiclassical Bloch-type equations
$$ \dot {\bf y}=\frac{1}{2}{\bf \bigtriangledown}{\cal H}^{cq}\times
{\bf \bigtriangledown}{\cal C},\; {\bf \bigtriangledown}{\cal H}^{cq}=
(2\Re\{g\}\Theta (y_0),- 2\Im\{g\}\Theta (y_0), a+2[\Re\{g\}y_1-
\Im\{g\}y_2]\partial \Theta (y_0)/\partial y_0), $$
$$\Theta (y_0)=\left(\frac{J\mp y_0}{2J}\right)^{\pm 2J-1}\sum_{f\geq 0}
\frac{(2J)^{-1}\,\sqrt{\Phi(\mp J+f)}\,[ y_0\pm J]^f}{f!(f+1)!N^2(J,f+1)
[J \mp y_0]^f}, \quad {\bf \bigtriangledown}{\cal C}=2(y_1,y_2,\pm y_0)
  \eqno (4.3)$$
At the same time, using substitutions (4.2) in Eqs. (3.6a) and inserting
them in Eqs. (4.1) one finds in the mean-field approximation (3.5),
respectively, canonical Hamiltonians equations
$$ \dot q = a\mp [\Re\{g\}\cos q+ \Im\{g\}\sin q]
[\pm (J^2-p^2)\Phi(p)]^{-1/2}\partial [(J^2-p^2)\Phi (p)]/\partial p, $$
$$\dot p = [-\Re\{g\}\sin q+ \Im\{g\}\cos q]\sqrt{\pm (J^2-p^2)\Phi(p)}
 \eqno (4.4a)$$
and more simple in comparison with Eqs. (4.3) nonlinear Bloch-type equations
obtained from Eqs. (4.3) by the substitution
$${\bf \bigtriangledown}{\cal H}^{cq}\rightarrow {\bf \bigtriangledown}
{\cal H}^{cmf} = $$
$$(2 \Re\{g\} [\Phi(y_0)]^{1/2},-2\Im\{g\} [\Phi(y_0)]^{1/2},
a+ [\Re\{g\} y_1- \Im\{g\}y_2)]
[\Phi(y_0)]^{-1/2}\partial \Phi (y_0)/\partial y_0)        \eqno (4.4b)$$
Note that these latter Bloch-type equations are equivalent to those obtained
in [10] in terms of variables $v_i(t)=<V_i(t)>$ and solved in terms of
hyperelliptic functions.

So, Eqs. (4.1)-(4.2) and their specifications (4.3)-(4.4) yield a tool
for examining quasiclassical dynamics of $SL(2)$ clusters within subspaces
$L([l_i])$. However, they are not suitable for such an analysis at
arbitrary initial conditions or for time-evolution of uncoupled
(characterizing single subsystems) dynamical variables that is often
necessary in practice. At the same time Eqs. (3.3) and (3.6) enable to obtain
appropriate QAs
$$U^{cq}_{H}(t)= \sum _{[l_i], v} S_Y(\xi_0)^{\dagger}
\; \exp (\frac{-i t E([l_i];v)}{\hbar}) \;|[l_i];v\rangle
\langle v;[l_i]| \;S_Y(\xi_0)=$$
$$ \sum _{[l_i]}\sum _{v\geq 0}\exp (\frac{-i t E([l_i];v)}{\hbar})
\sum_{f\geq 0}\sum_{f'\geq 0} S_{f v}(J;g,r)S^*_{f' v}(J;g,r)
|[l_i];f\rangle \langle f';[l_i]|
    \eqno (4.5)$$
of evolution operators $U_{H}(t)$ when eigenenergies $E([l_i];v)$ are
given by Eqs. (3.3a)-(3.4) or Eqs. (3.6). Evidently, $L([l_i])$-restrictions
$P_{[l_i]} U^{cq}_{H}(t)$ of such evolution operators (4.5) are distinguished
from evolution operators $U^{l}_{H;[l_i]}(t)$ associated with solutions of
Eqs. (4.1a).

Substitutions of Eqs. (3.3a)-(3.4) or Eqs. (3.6) in Eqs. (4.5) enable us to
calculate appropriate QAs for time-dependences
$$<F(t)>= \mbox{Tr}[U^{cq}_{H}(t)\,\rho\, U^{cq}_{H}(t)\,F]  \eqno (4.6)$$
of any dynamical variables $F$ where $\rho$ is a density operator for an
initial quantum state. For example, inserting in Eq. (4.6) ordered
exponentials of coupled ($Y_i$) or uncoupled (original) dynamical variables
one may get (after an appropriate Fourier transformation) formulas describing
dynamics of different (associated with GCS of both $SL(2)$ and dynamic
symmetry groups of subsystems) types of $Q$-,$P$- and Wigner quasiprobabilty
functions which are widely used for visualizing features of systems under
study [14,26]. Note also that, due to Eqs. (2.5), the
first line in Eq. (4.5) is more suitable for using Eq. (4.6) with $F=
F(\{Y_i\})$ whereas the second one is more relevant for calculations
with $F$ depending on uncoupled dynamical variables.

\section{Applications to a class of quantum-optical models}

In this Section we manifest a physical meaning of general results above on
ceveral concrete models which are widely applied in quantum optics, laser
physics and quantum electronics [3,4,8,18,19,25]. Specifically, as was shown
in [11], a natural area of applications of the $sl_{pd}(2)$ formalism is
provided by quantum models with Hamiltonians
$$ H_1/\hbar = \sum _{i=1}^2 \omega_i a_i^+ a_i  +  g' (a^+_1)^m (a_2)^n +
g'^* (a_1)^m (a^+_2)^n, \;n\leq m,
\eqno (5.1a)$$
$$ H_2/\hbar = \sum _{i=1}^m \omega_i a_i^+ a_i  + \omega_0 a_0^+ a_0  +
 g' (a^+_1...a^+_m) (a_0)^n + g'^* (a_1...a_m) (a^+_0)^n,  \;n\leq m,
\eqno (5.1b)$$
$$ H_3/\hbar = \omega_1 a_1^+ a_1 + \sum _{i=1}^N  [\sigma_0 (i) \epsilon/2+
g' \sigma_{+} (i)(a _1)^n + g'^* \sigma_{-} (i)(a _1^+)^n ]
 \eqno (5.1c)$$
where $g'$ are coupling constants, $a_{i}, a _{i}^+$ are boson operators
describing field modes with frequencies $\omega_i$, $\sigma_{\alpha}(i)$ are
Pauli matrices, $\epsilon$ is an energy difference of two level atoms and
non-quadratic parts of $H_i$ describe different multiphoton processes of
scattering and frequency conversion (Eqs. (5.1a,b)) as well as the
matter-radiation interactions in $n$-photon point-like Dicke models in
rotating wave approximation (Eqs. (5.1c)). Note that in applications, one
considers, as a rule, models (5.1) with $n=0,1$ that correspond,
respectively, to semiclassical or completely quantum versions of models under
study [3,4,8,18,23,25].

Appropriate Hilbert spaces $L(H_i)$ are multimode Fock spaces
$L_F=Span\{|\{n_i\}\rangle=\prod_i [n_i!]^{-1/2}
(a_i^+)^{n_i}|0>\}$ for models (5.1a)-(5.1b) whereas for models (5.1c)
$L(H_3)$ are direct products of single-mode Fock spaces $L_F$ and "atom"
spaces $L_a=Span\{|j,\mu;\{j_{int}\}>\rangle\}$ where $|j,\mu;\{j_{int}\}>$
are the basis vectors of irreducible representations of the "atom" group
$SU(2)^a$ (with generators $\Sigma_{\alpha}=\sum_{i=1}^N \sigma_{\alpha}(i)$)
which are obtained from one-atom basis states $|\pm>(i)$ with the help of the
generalized Wigner coefficients and $\{j_{int}\}$ are sets of the $SU(2)^a$
intermediate angular momenta labeling basis vectors of the irreducible
representations of the symmetric group $S_N$ and being integrals of
motion [11].

Hamiltonians (5.1) are expressed in the form (1.1)-(1.2) with the help of
introducing $sl_{pd}(2)$ dynamic variables $V_0,V_+,V_-=(V_+)^{\dagger}$ and
integrals of motion $R_j$ via a generalized Jordan-Schwinger mapping [10]
given for $H_1,H_2,H_3$ respectively as follows [11]:
$$V_0=\frac{1}{m+n} (a^+_1 a_1-a_2^+a_2),\; V_+ = (a_1^+ )^m(a_2)^n,
\quad R_1 = \frac{1}{m+n}(na^{+}_1 a_1+ma_2^+a_2),
\eqno (5.2a)$$
$$V_{0} =\frac{1}{m+n}(\sum_{i=1}^{m}a^{+}_{i} a_{i} - a_{0}^{+}a_{0}),
\qquad V_{+} = a_{1}^{+}...a_{m}^{+}(a_{0})^{n},$$
$$R_k=\frac{1}{m+n}(a^{+}_j a_j - a_{j+1}^+a_{j+1}),\, k=1,\dots ,m-1,
\quad R_m =\frac{1}{m+n}(n \sum_{i=1}^{m}a^+_i a_i + m a_0^+a_0),
\eqno (5.2b)$$
$$V_{0} =\frac{1}{2} \sum_{i=1}^{N}\sigma_{0}(i), \quad
V_{+} = \sum_{i=1}^{N} \sigma_{+}(i)(a_{1})^{n},
\quad R_1= \frac{n}{2}\sum_{i=1}^{N} \sigma_{0}(i)+a_{1}^+a_{1}
\eqno (5.2c)$$

The structure polynomials $\Psi (V_0)$ are determined with the help of Eqs.
(5.2) (and defining relations for $a(i), a^+(i), \sigma_{\alpha}(i)$) from
Eq. (1.3) which is valid for all $L(H_i)$. Then for $H_1,H_2,H_3$ one finds,
respectively,
$$\Psi(V_{0})=(m V_0 + R_1)^{(m)} (R_1-nV_0 + n)^{(n)},
 \eqno (5.3a)$$
$$\Psi(V_{0})=[R_m -nV_0 +n]^{(n)}([R_m -(m+n)\sum_{i=1}^{m-1}i R_i]/m +V_0)
\,N_1\, N_2\,\dots\, N_{m-1},$$
$$ N_k=\frac{1}{m}[R_m -(m+n)\sum_{i=1}^{m-1}i R_i] +V_0 +
(m+n)\sum_{i=k}^{m-1} R_i,\, k=1,\dots ,m-1,
\eqno (5.3b)$$  
$$\Psi(V_{0})= [C_{2}(2) - V_{0}^{(2)}][R_1-nV_0 +n]^{(n)}   
                                  \eqno (5.3c)$$
where $C_{2}(2)=\Sigma_+ \Sigma_- +(\Sigma_0/2)^{(2)}$ is the Casimir
operator of the "atom" $su(2)$ algebra and $[C_{2}(2), V_{\alpha}]=0,
A^{(B)} =A(A-1)...(A-B+1)$.

The subspaces $L([l_i])$ in Eq. (1.4) are generated by the lowest vectors
$|[l_i]>$ which are given for different $H_i$ as follows\\
$H_1$ :
$$ |[l_i]>=|\{n_1=\kappa,n_2=s\}\rangle,\quad l_0 =
\frac{1}{m+n}(\kappa-s), \quad l_{1} =\frac{1}{m+n} (n\kappa+ms),$$
$$R_j |[l_i]> = l_j |[l_i ]>,\;\kappa= 0,1,\dots,m-1,\;   s= 0,1,...,
\eqno (5.4a)$$
$H_2$ :
$$\quad |[l_i]>=|\{n_1=\kappa_1,n_2=\kappa_2,\dots,n_m=\kappa_m,
n_0=s\}\rangle,\qquad \prod_i^m \kappa _i=0,$$
$$ l_{0}=\frac{1}{m+n}(\sum_{i=1}^m\kappa_i-s),\; l_{m} =
\frac{1}{m+n}(n\sum_{i=1}^{m}\kappa_i + m s),\;l_k =\frac{(\kappa_j -
\kappa_{j+1})}{m+n},\, k=1,\dots,m-1,$$
$$ R_j |[l_i]> =l_{j} |[l_{i}]>,\;\kappa_i= 0,1,...,\; s= 0,1,...,
\eqno (5.4b)$$
$H_3$ :
$$ \qquad\qquad |[l_i]>=|\{n_1=\kappa\}\rangle |j,\mu =
-j;\{j_{int}\}>,
\quad l_0 = -j,\quad  l_1 = \kappa -n j,$$
$$R_j |[l_{i} ]> = l_{j} |[l_i]>, \quad \kappa= 0,1,...,\; 0 \leq j \leq N/2,
   \eqno (5.4c)$$
where $|\{n_i\}\rangle$ are standard Fock states and $|j,-j;\{j_{int}\}>$
are lowest vectors of irreducible representations of the "atom" group
$SU(2)^a$. From Eqs. (5.3)-(5.4) it follows that we have compact versions
of algebras $sl_{pd}(2)$ in all cases except for models (5.1a,b) with
$n=0$.

Eqs. (5.3)-(5.4) yield requisites for specifications of general results of
Sections 3 and 4. However, for the sake of simplicity of our exposition, we
restrict ourselves by considering certain simple examples which elucidate
main features of new QAs and, simultaneously, will provide a base for
further investigations of the most spread in applications models and
physically important cases ($n\leq 1, m\leq 3$ in Eqs. (5.1a)-(5.1b)
and $n=1$ in Eq. (5.1c)).

{\it Example 1. "Cluster" mean-field energy spectra in models (5.1) with
$n=1, m= 2,3$.} Inserting Eqs. (5.3) in compact ($su(2)$) versions of Eq.
(2.1a) and using Eqs. (5.4) one finds for the  polynomials $\Phi (Y_0)$
from Eqs. (2.1a) the following expressions
$$H_1 :\;\Phi^m (Y_0)\;=\;\frac{(mY_0+mJ+m+\kappa)^{(m)}}
{(J+1+Y_0)},\qquad\;J=s/2,\qquad\;X^{(m)}=X(X-1)\dots,
 \eqno (5.5a)$$
$$H_2 :\qquad\Phi^m (Y_0)\;=\;\prod_{i=1}^{m}{'}(Y_0+J+\kappa_i+1)\;=
\;\prod_{i=1}^{m}{'}(Y_0+\frac{s}{2}+\kappa_i+1),
\qquad\prod_{i=1}^{m}\kappa_i =0,
\eqno (5.5b)$$
$$H_3 :\; \Phi (Y_0)= \max(\kappa,2j)-J-Y_0 =
\left\{
\begin{array}{rcl}
(\kappa-j- Y_0), \quad \kappa\geq 2j,\\
(2j-\frac{\kappa}{2}-Y_0), \quad \kappa\leq 2j\\
\end{array}
\right.
,\, J=\min(j,\frac{\kappa}{2})
 \eqno (5.5c)$$
where it is also taken into account that $d([l_i])=2J+1$ and simultaneously
$d([l_i])=s+1$ for $H_i, i=1,2$ and $d([l_i])=\min(2j,\kappa)+1$ for $H_3$;
besides, the numerator in Eq.(5.5a) always contains the factor
$(J+1+Y_0)$ due to the definition of the symbolic powers $X^{(m)}$
and $\prod_{i=1}^{m}{'}$ in Eq. (5.5b) means that in the product the
term with $\kappa_i=0$ is omitted.
Then, inserting Eqs.(5.5) in Eqs. (3.3),(3.4),(3.6) and Eqs. (4.3)-(4.5) and
using  also Eqs. (5.4) one can obtain appropriate specifications of QAs above
for energy spectra, Bloch-type dynamical equations and evolutions operators
and to examine their features depending on characteristic parameters of
models under study.

However, postponing such a detailed analysis for further publications, we
only write down appropriate specifications of Eqs. (3.6a) in
the resonace cases ($a=0$ in Eq. (2.2)) and at $m=2$ in Eqs. (5.5a,b) when
Eqs. (3.6b) are solved analytically yielding
$$\cos 2r=c(s,\kappa)=\frac{1}{3}\left (\frac{2\kappa+1}{s}+1-2\sqrt{1+
(\frac{2\kappa+1}{2s})(\frac{2\kappa+1}{2s}+1)}\right )$$
$$ \approx\frac{1}{3}(\frac{2\kappa+1}{2s}-1) (\mbox{for}\,s > 2\kappa+1)
  \eqno (5.6a)$$
$$\cos 2r=c(s,\kappa)=\frac{1}{3}\left (\frac{2\kappa+2}{s}+1-2\sqrt{1+
(\frac{\kappa+1}{s})(\frac{\kappa+1}{s}+1)}\right )$$
$$\approx\frac{1}{3}(\frac{\kappa+1}{s}-1) (\mbox{for}\,s > \kappa+1)
\eqno (5.6b)$$
$$\cos 2r= c(j,\kappa)=\frac{1}{3}\left (1-2\mu(\kappa,j)+2\sqrt{1-
\mu(\kappa,j)+(\mu(\kappa,j))^2}\right ),
\quad \mu(\kappa,j)=\frac{\max(\kappa,2j)}{\min(\kappa,2j)}  \eqno (5.6c)$$
for $H_1,H_2,H_3$ respectively. Then, with the help of Eqs. (5.5)-(5.6)
eigenenergies $E^{cmf}([l_i];v)$ in the "cluster" mean-field approximation
(3.5) are given as follows,\\
$H_1$:\\
$$E^{cmf}([l_i];v)/\hbar-\tilde C=E^{cmf}(\kappa, s; v)/\hbar
-(\kappa +2s)\omega_1= $$
$$-|g'|(s -2v)\sin 2r\sqrt{2((-s +2v)\cos 2r+s+ 2\kappa+1)}=$$
$$-|g'|(s -2v)\sqrt{(1-[c(s,\kappa)]^2)2((-s+2v)c(s,\kappa)+s+
2\kappa+1)},\; \omega_2 =2 \omega_1
  \eqno (5.7a)$$
$H_2$:\\
$$E^{cmf}([l_i];v)/\hbar-\tilde C=E^{cmf}(\kappa, s; v)/\hbar
-\kappa_1\omega_1+\kappa_2\omega_2 +s(\omega_1+\omega_2) = $$
$$-|g'|(s -2v)\sin 2r\sqrt{(-\frac s 2+v)c(s,\kappa)+\frac s 2+\kappa+1)}=$$
$$-|g'|(s -2v)\sqrt{(1-[c(s,\kappa)]^2)((-\frac s 2+v)c(s,\kappa)+
\frac s 2+\kappa+1)},$$
$$\kappa_1\kappa_2=0,\quad\kappa=\max(\kappa_1,\kappa_2),
\quad\omega_0=\omega_1+\omega_2
  \eqno (5.7b)$$
$H_3$:\\
$$E^{cmf}([l_i];v)/\hbar-\tilde C=E^{cmf}(\kappa,j;v)/\hbar-(\kappa-
j)\omega_1=$$
$$-|g'|(s -2v)\sin 2r\sqrt{\max(\kappa,2j)-J-(-J+v)\cos 2r}=$$
$$-|g'|(s -2v)\sqrt{(1-[c(s,\kappa)]^2)(\max(\kappa,2j)-J-(-J+v)
c(s,\kappa))},\quad \epsilon= \omega_1,\; J=\min(j,\frac{\kappa}{2})
  \eqno (5.7c)$$

Evidently, Eqs. (5.7) manifest explicitly an essentially nonlinear dependence
of energy levels $E^{cmf}([l_i];v)$ on their both $su(2)$-invariant ($\kappa,
s,j$) and non-invariant ($v$) labels unlike standard QAs obtained by means of
using in Eqs. (3.1) GCS associated with dynamic symmetry algebras of
subsystems. Indeed, using in variational schemes (3.1)-(3.2) Glauber's CS
$\prod_i{\cal D}(\alpha_i)|\{n_i\}\rangle,
\,{\cal D}(\alpha_i)= \exp (\alpha_i a^+_i -\alpha_i^* a_i)$ for models (5.1)
and, additionally, "atomic" $SU(2)^a$ GCS $\exp(\xi \Sigma_+ -\xi^*\Sigma_-)
|j,\mu;\{j_{int}\}>$ for models (5.1c) as trial functions ,
one finds the following analogs of Eqs. (5.7) for such simplest QAs
$$H_1:\quad \qquad E^{smf}(n_1,n_2)/\hbar\;=\;  \omega_1(n_1 + 2n_2)+
\Lambda_1 (\omega_1,|g'|),\qquad \omega_2 =2 \omega_1  \eqno (5.8a)$$
$$H_2:\;E^{smf}(n_1,n_2,n_0)/\hbar=\omega_1(n_1 + n_0)+\omega_2(n_2 + n_0)+
\Lambda_2 (\omega_1,\omega_2,|g'|),\;\omega_0= \omega_2 + \omega_1
  \eqno (5.8b)$$
$$H_3:\quad \qquad E^{smf}(n_1,\mu)/\hbar\;=\;\omega_1 n_1+ 
\mu\Omega (\omega_1,|g'|) + \Lambda_3 (\omega_1,|g'|), \qquad \epsilon= 
\omega_1  \eqno (5.8c)$$
where $\Lambda_i(\dots)$ are constant (for whole $L(H)$) energy shifts and
$\Omega$ is an efficient frequency.
Evidently, energy levels (5.8) depend linearly on $G_{inv}$-noninvariant
labels $n_i$ arranged on multidimensional lattices that provides
multiperiodic dynamical regimes. Other ordinary QAs [4,8,25], e.g., obtained
with the help of GCS of partially coupled subsystems, lead to similar
results (as it is seen, in fact, from comparisons of Eqs.(5.8b) and (5.8c)).

{\it Example 2. "Cluster" mean-field energy spectra in models (5.1a,b) with
$n=0, m=3$.} In this case models under study yield so-called parametric
approximations for models of the first example with $m=3$. Besides, we have
noncompact versions of $sl_{pd}(2)$ algebras because all subspaces $L([l_i])$
are infinite-dimensional. This, in turn, causes an ambiguity of determining
the parameter $J$ in the generalized Holstein-Primakoff mappings (2.1) on
subspaces $L([l_i])$ because a polynomial character of $\Phi (Y_0)=
\Psi (Y_0+l_0- J+1)/(J+ Y_0)(- J+1+Y_0)$ is provided by two values of $J$ on
each subspace $L([l_i])$ that requires to add a choice procedure of $J$ to
Eqs. (3.1). However, we restrict ourselves by writing down analogs of
Eqs. (5.5a,b),
$$H_1 :\qquad\Phi^3 (Y_0)\;=\;\frac{(3Y_0-3J+3+\kappa)^{(3)}}{(-J+1+Y_0)
(J+Y_0)}=27(Y_0 +\lambda_1 (\kappa, J_{\kappa})),\qquad \kappa =0,1,2,
 \eqno (5.9a)$$
$$H_2 :\qquad\Phi^3 (Y_0)\;=\; \frac{\prod_{i=1}^3 (Y_0-J+1+\kappa_i)}
{(-J+1+Y_0)(J+Y_0)} =Y_0 +\lambda_2 (\{\kappa_i\}, J_{\kappa_i}),
\qquad\prod_{i=1}^{3}\kappa_i =0,
\eqno (5.9b)$$
which, nevertheless, manifest differences of parametric QAs from those given
by Eqs. (5.5) due to linear and quadratic forms of $\Phi^3 (Y_0)$ in these
cases (constants $\lambda_i(\dots, J_{\dots})$ are easily determined for
chosen values $\{\kappa_i\},J_{\kappa_i}$). A more detailed analysis of such
comparisons will be given elsewhere.

\section{Conclusion}

So, we have obtained new approximations for energy spectra and evolution
operators as well as nonlinear Bloch-type dynamic equations for models (1.1)
(and (5.1)) by means of using the mapping (2.1) and standard variational
schemes [17,5] with the $SL(2)$ GCS as trial functions. They may be called as
"cluster" (or correlated) QAs owing to taking into account strong quantum
correlations between interacting subsystems. These approximations may be used
to calculate in models of the (5.1) type time evolution of different
quantum-statistical characteristics and quasidistributions (cf. [8,14])
and to find bifurcation sets and solutions of nonlinear Hamiltonian
flows determined by Eqs. (4.1) and (4.5) (cf. [5,23]). In this way
we hope to reveal in these models new cooperative phenomena
and dynamical regimes (due to quantum correlations between subsystems)
by analogy with those found in [9,18,22,23,27] and many other papers
by using standard QAs; herewith different QAs above are expected to
elucidate the role of such correlations depending on a choice of
initial quantum states and paprameters of models under study (cf. [22,23]).

However, from the practical point of view for this aim it is desirable to
modify and to simplify Eqs. (3.3) and (4.5) by using different properties of
the hypergeometric functions ${_2F_1(a,b;c;x)}$, including their integral
representations and asymptotic expansions [20,28]. (Specifically, in such a
way one can express spectral functions $E^{\Phi}_f(r;J;v)$ in terms of the
hypergeometric functions ${_4F_3(...;1)}$ which are proportional to the
$sl(2)$ Racah coefficients [12].) Along this line it is also
of importance to get estimations of accuracy of QAs obtained and of their
efficiency in comparison with other approximations (e.g., considerd in
[8,9,10,18]). One way to  do such estimations is in comparisons of these QAs
with appropriate computer calculations (cf. [18,27]) and another one is
connected with using the "energy-trace" proximity meausures (3.7).

Another line of further investigations concerns developments of mathematical
aspects of the work. Indeed, results of Sections 3,4 correspond to picking
out "smooth" $sl(2)$ factors $S_Y(\xi_0)=\exp(\xi_0 Y_+-\xi_0^* Y_-)$ in
exact (generally, not "smooth") diagonalizing operators $S$ determined by
Eqs. (2.9), (2.12) and, when using Eqs. (4.1a), in evolution operators
$U_H(t)$ determined by exact evolution equations given in [11]; besides,
Eqs. (4.5) yield another type of QA for evolution operators $U_H(t)$. All
these QAs can be used as initial approximations in iterative schemes of
constructing exact solutions which are similar to those developed to examine
nonlinear problems of classical mechanics and optics [29] or in search of
suitable multi-parametric improvements of variational schemes used by
introducing "form-factors" with extra fitting parameters in original trial
functions. It is also of interest to develop methods of obtaining simple
formulas for exact solutions of tasks under consideration in order to compare
with them results of approximations found above. At present one may to point
out, at least, three promising ways along this line.

One of them is in simplifications of integral solutions obtained in [11] for
both  evolution and spectral tasks. The second way, leading to  solving
singular differential equations, is connected with using
two conjugate differential realizations of $sl_{pd}(2)$ generators
$V_{\alpha}$ [10,12]:
$$ V_+=z,\quad V_0=zd/dz+l_0,\quad V_-= z^{-1} \Psi(zd/dz+l_0),
\eqno (6.1a)$$
$$ V_-=d/dz,\quad V_0=zd/dz+l_0,\quad V_+=\Psi(zd/dz+l_0)(d/dz)^{-1}
\eqno (6.1b)$$
which are, in turn, related to realizations of $sl_{pd}(2)$ generators
$V_{\alpha}$ by quadratic forms in $sl(2)$ generators $Y_{\alpha}$ taken in
the coherent-state representations (cf. [30,15]).
(In fact, these realizations
were used implicitly when obtaining exact integral solutioms [11].)
For example, when the structure polynomial $\Psi (x)$ has the third degree
(as, e.g., in models (5.1) with $n=1,m=2$), the realization (6.1b) reduces
original tasks to solutions of the Riccati equations [12]. In this
connection one may consider the hypergeometric functions ${_2F_1(a,b;c;x)}$
determining QAs obtained as specific asymptotics of new classes of special
functions determining exact solutions that opens a possibility to use the
techniqe of asymptotic expansions [28] for finding latters. Last (but not
least!) way is due to interrelationships between $sl_{pd}(2)$ algebras and
certain $q$-deformed algebras mentioned in Section 2 that enables us to use
for purposes formulated above techniques of $q$-deformed algebras and
$q$-special functions, in particular, $q$-exponents defined with the help of
the coherent states map of the paper [21]. Evidently, a progress in solving
all these problems will promote to an extension of the orbit type GCS concept
[16] and, simultaneously, to a more fine description of "classical" phase
spaces associated with dynamic symmetry algebras $sl_{pd}(2)$ (cf. [31]).

The work along these lines is now in progress.

\section{Acknowledgements} 

Preliminary results of the work were reported at the VII International
Conference on Symmetry in Physics (JINR, Dubna, July 10-16, 1995), at
the XV-XVI Workshops on Geometric Methods in Physics (Bialowieza, Poland,
July 1-7, 1996 and June 30-July 6, 1997) and on the Seminar of
Arbeitsgruppe "Nichtclassishe Strahlung" der Max-Planck-Gesellschaft an
der Humboldt-Universitaet zu Berlin (Berlin, December 9, 1996). The author
thanks  Professors S.T. Ali,  A. Odzijewicz and H. Paul
and Doctors S.M. Chumakov, C. Daskaloyannis, A. Bandilla and A. Wuensche for
useful discussions and remarks. The paper is prepared under a partial
support of the Russian Foundation for Basic Research, grant No 96-02 18746a.

\end{document}